\documentclass[conference]{IEEEtran}
\IEEEoverridecommandlockouts

\usepackage{cite}
\usepackage{amsmath,amssymb,amsfonts}
\usepackage{algorithmic}
\usepackage{graphicx}
\usepackage{textcomp}
\usepackage{xcolor}

\usepackage{hyperref}
\hypersetup{
colorlinks=true,
linkcolor=black
}

\usepackage{subfigure}
\def\BibTeX{{\rm B\kern-.05em{\sc i\kern-.025em b}\kern-.08em
    T\kern-.1667em\lower.7ex\hbox{E}\kern-.125emX}}
\begin{document}

\title{
Exploring Differences between Human Perception and Model Inference in Audio Event Recognition
}
\author{
\IEEEauthorblockN{Yizhou Tan\IEEEauthorrefmark{1}, Yanru Wu\IEEEauthorrefmark{1}, Yuanbo Hou\IEEEauthorrefmark{2}, Xin Xu\IEEEauthorrefmark{3}, Hui Bu\IEEEauthorrefmark{3} \\
Shengchen Li\IEEEauthorrefmark{1}, Dick Botteldooren\IEEEauthorrefmark{2}, Mark D. Plumbley\IEEEauthorrefmark{4}}
\\
\IEEEauthorrefmark{1}Xi'an Jiaotong-Liverpool University, Suzhou, China \\
\IEEEauthorrefmark{2}WAVES Research Group, Ghent University, Belgium 
\IEEEauthorrefmark{3}Beijing AISHELL Technology Co., Ltd, China \\
\IEEEauthorrefmark{4}University of Surrey, Guildford, United Kingdom \\
\{Yizhou.Tan22, Yanru.Wu21\}@student.xjtlu.edu.cn, Shengchen.Li@xjtlu.edu.cn \\
\{Yuanbo.Hou, Dick.Botteldooren\}@UGent.be, 
\{Xuxin, Buhui\}@aishelldata.com,  
M.Plumbley@Surrey.ac.uk}


\maketitle

\begin{abstract}
Audio Event Recognition (AER) traditionally focuses on detecting and identifying audio events. Most existing AER models tend to detect all potential events without considering their varying significance across different contexts. This makes the AER results detected by existing models often have a large discrepancy with human auditory perception. Although this is a critical and significant issue, it has not been extensively studied by the Detection and Classification of Sound Scenes and Events (DCASE) community because solving it is time-consuming and labour-intensive. To address this issue, this paper introduces the concept of semantic importance in AER, focusing on exploring the differences between human perception and model inference. This paper constructs a Multi-Annotated Foreground Audio Event Recognition (MAFAR) dataset, which comprises audio recordings labelled by 10 professional annotators. Through labelling frequency and variance, the MAFAR dataset facilitates the quantification of semantic importance and analysis of human perception. By comparing human annotations with the predictions of ensemble pre-trained models, this paper uncovers a significant gap between human perception and model inference in both semantic identification and existence detection of audio events. Experimental results reveal that human perception tends to ignore subtle or trivial events in the event semantic identification, while model inference is easily affected by events with noises. Meanwhile, in event existence detection, models are usually more sensitive than humans.
\end{abstract}


\section{Introduction}
Audio Event Recognition (AER) is vital for detecting and identifying audio events within an audio stream. With the advancement of AER research \cite{chan2020comprehensive,mesaros2021sound,tan2023transductive, hou2023ai}, a large volume of labelled audio data \cite{audioset, DCASE_Task5_Dataset,mesaros2016tut,politis2020dataset, piczak2015esc, fonseca2019learning} has been generated to build systems capable of detecting all potential audio events. However, the importance of specific audio events can vary significantly depending on the context in which they occur \cite{oliva2007role}. For example, the sound of a running car might be considered background noise in a bustling city, but the same sound in a remote forest could signify a rare human presence or an unusual occurrence. 
Given that human perception naturally assigns different levels of attention to the same event in varying contexts, AER systems need to go beyond mere detection and incorporate an understanding of the semantic importance of audio events based on their environmental relevance.
This work aims to construct the notion of semantic importance in AER, drawing on human understanding to analyze the gap between current model inference and human perception.

Although several works have verified the scene of audio is identifiable \cite{abesser2020review,cai2024leveraging,cai2024tf,tan_asc, 10274856} and introduced the context information to improve the audio-pattern recognition performance \cite{heittola2013context}\cite{tonami2022sound} \cite{chang2021context}, there are few researches to explore how to quantify the semantic importance. This work assumes that the semantic importance of the same events may vary among individuals or groups. Due to cognitive and cultural differences, individuals may focus on specific events to varying degrees, indicating that the sole annotation of ground truth can not quantitatively reflect semantic importance.

To solve the data annotation problem, this work constructs an available Multi-Annotated Foreground Audio Event Recognition (MAFAR) dataset containing 180 minutes of audio data and 10 different sets of labels.
The dataset was collected during a real trip in 4 cities, including Shanghai, London, Gent, and Rotterdam, by a GoPro device. During the annotation process, each audio file was independently labelled by 10 professional annotators. Unlike conventional datasets with predefined labels, this work did not provide a closed label set in advance. Instead, annotators were asked to label ``prominent and noticeable" events as foreground events using descriptive texts to reflect their original perception. In the later experiments, these descriptive texts were aligned to a subset (86 classes) of the large-scale AudioSet \cite{audioset} by a large language model GPT-4 \cite{gpt4}, but the alignment is not compulsory for future works.

Based on the MAFAR dataset, this work tries to use the variance and frequency in the annotation distribution to define the perception of the human group in two dimensions: event semantic identification and existence detection. The variance reflects the level of agreement among annotators regarding event identification, while the frequency indicates the interest weight of the existing events. Based on this definition, this work explores the differences between human perception and model inference in two key areas: event semantic identification and event existence detection. 

By comparing the variance of human annotation and ensemble model inference, the experiments uncover humans tend to overlook subtle or trivial events in event semantic identification, while models are more prone to being influenced by noisy events. In event existence detection, by adjusting annotation frequency thresholds to set the different ground truths, the changes in model performance highlight that models have greater sensitivity than humans, effectively identifying events even at lower annotation frequencies.


This paper is organized as follows: 
Section II introduces the MAFAR dataset.
Section III shows the benchmark method and results of measuring the differences between human perception and model inference.
Section IV discusses the potential reason for the results.
Section V concludes our work.

\section{MAFAR Dataset}\label{section2}
\begin{figure}[t]
  \centering
  \includegraphics[width=0.48\textwidth]{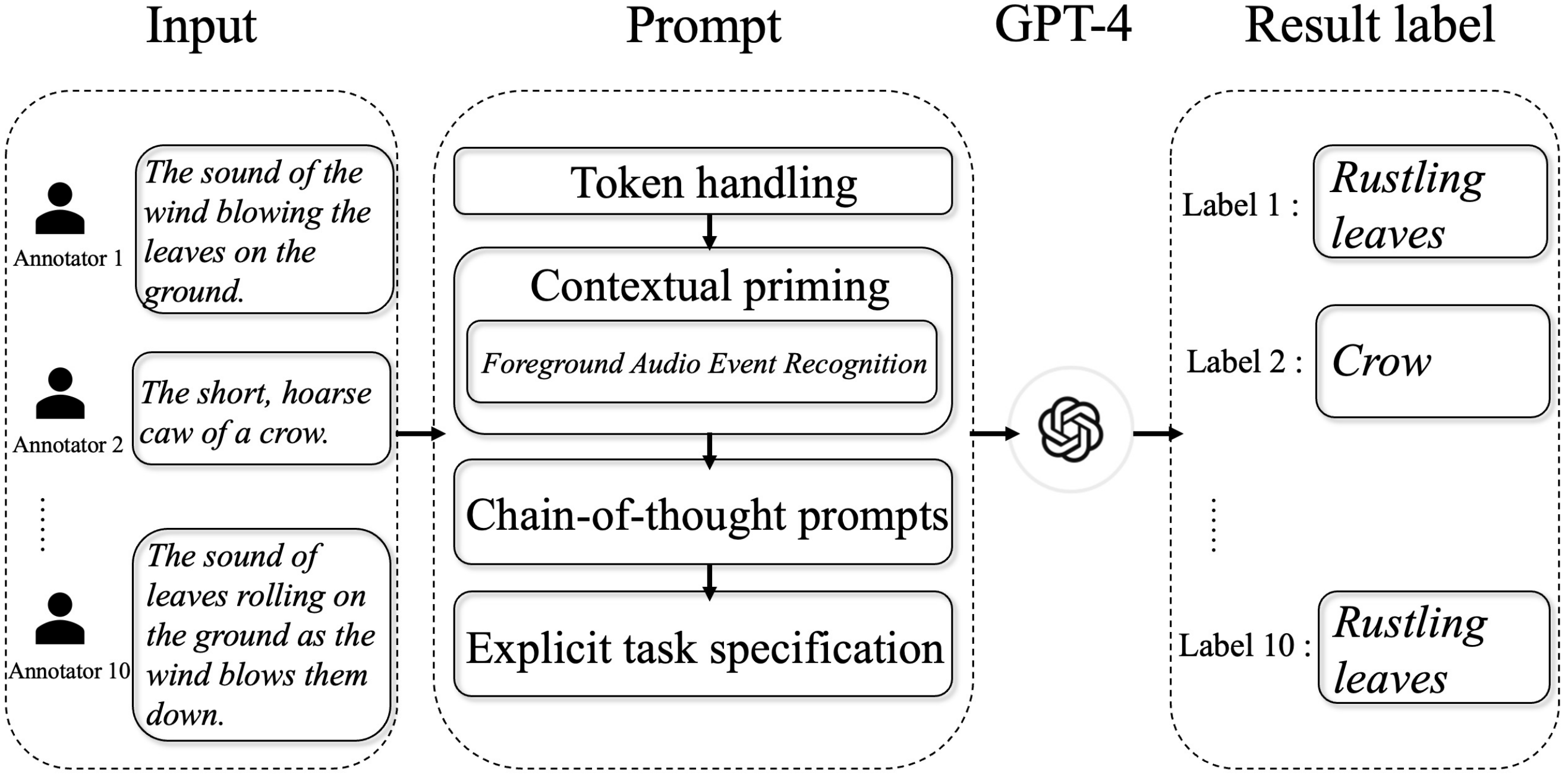}
  \caption{The process of label alignment is assisted by GPT-4.  
  }
  \label{fig:gpt4}
\end{figure}
All data from the MAFAR dataset was collected from a trip in Shanghai, London, Gent, and Rotterdam by a GoPro device, while the GoPro was worn on the chest of a person walking through these cities. The raw data is stored in video format, consisting of 25 files and 180 minutes in total. For each file, 10 professional annotators are required to annotate the ``prominent and noticeable'' events in the audio, mainly referring to the audio information and supplemented by visual information. The annotation comprises the file name, start time, end time and a descriptive text of the labelled period. (Dataset homepage: \url{https://github.com/Voltmeter00/MAFAR})

\subsection{Annotation Guidelines for Humans}
Since there is no pre-given set of categories, the whole annotation process mainly relies on human perceptions and the following guidelines:
\begin{enumerate}
\item Foreground sounds are distinctly prominent and noticeable in the scene while watching the video or listening to the audio. These sounds are crucial to the scene and significantly contribute to the overall experience.

\item Annotate the timestamps for the start and end of the foreground sound. The duration is not restricted, and the focus should be on the completeness of the event.

\item Use simple words to describe the audio events, adding adjectives like ``close," ``loud," ``sharp," etc. 

\end{enumerate}

\subsection{Label Alignment via Large Language Model}
To facilitate the involvement of descriptive text labels in training deep learning models, this work tries to align human annotations with a commonly used label set, namely 527 classes of AudioSet \cite{audioset}. 
The challenges were encountered in translating free-form natural language descriptions into predefined categories. Descriptive and subjective terms were often used by human annotators to label the audio events, making the direct mapping to a fixed set of categories complex. 
A large language model (LLM), specifically GPT-4 \cite{gpt4}, was utilized to assist in converting human-generated labels into AudioSet labels, with the following pipeline: 
\begin{enumerate}
\item 
The GPT-4 tried to align the provided descriptive texts to the most relevant AudioSet class, as shown in Fig. \ref{fig:gpt4}. 

\begin{figure}[t]
  \centering
  \setlength{\abovecaptionskip}{0.1cm}  
	\setlength{\belowcaptionskip}{-0cm}
  \includegraphics[width=0.45\textwidth]{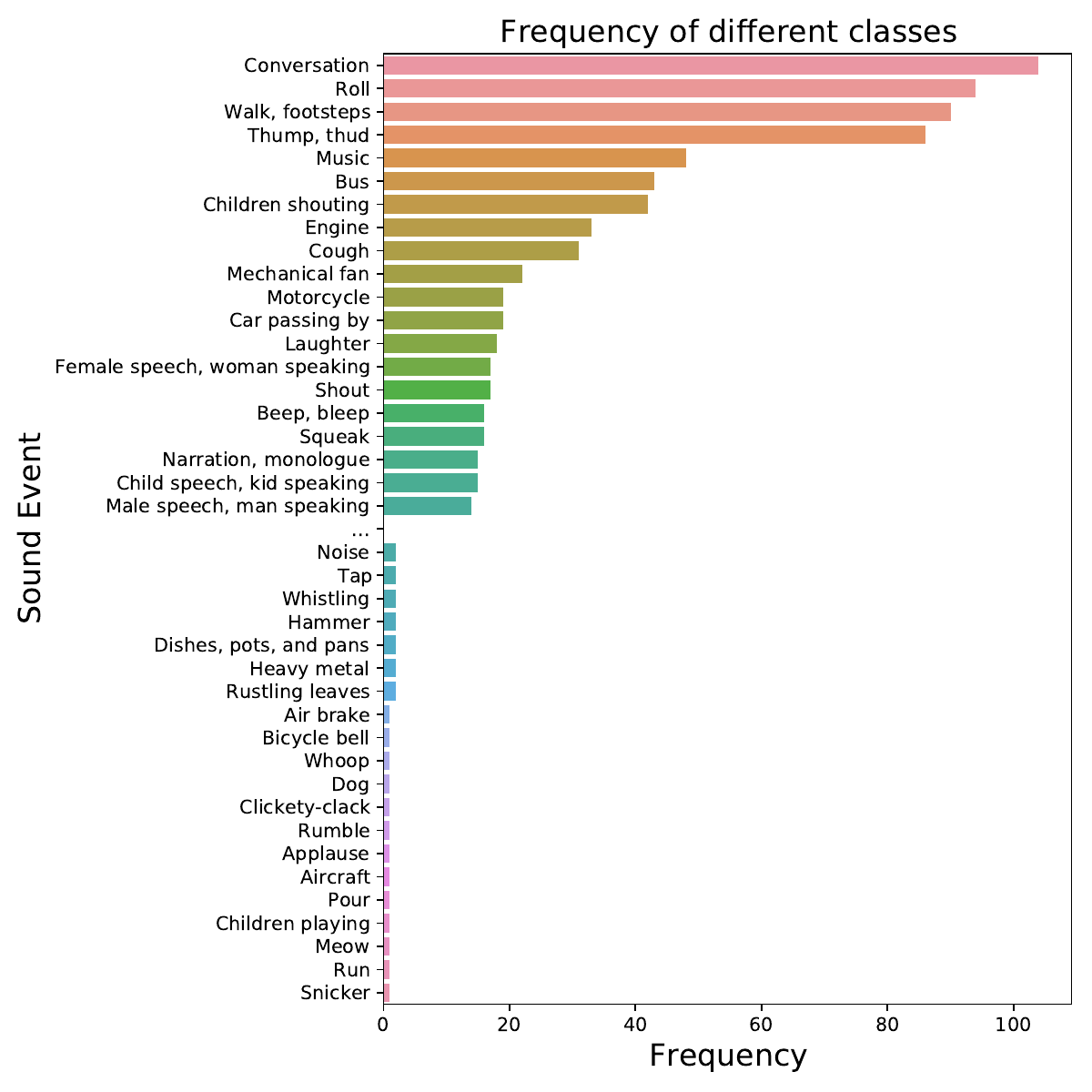}
  \caption{The frequency of 86 classes of audio events after label alignment.}
  \label{fig:frequency}
\end{figure}

\begin{figure*}[t]
  \centering
  \includegraphics[width=0.85\textwidth,height=10.5cm]{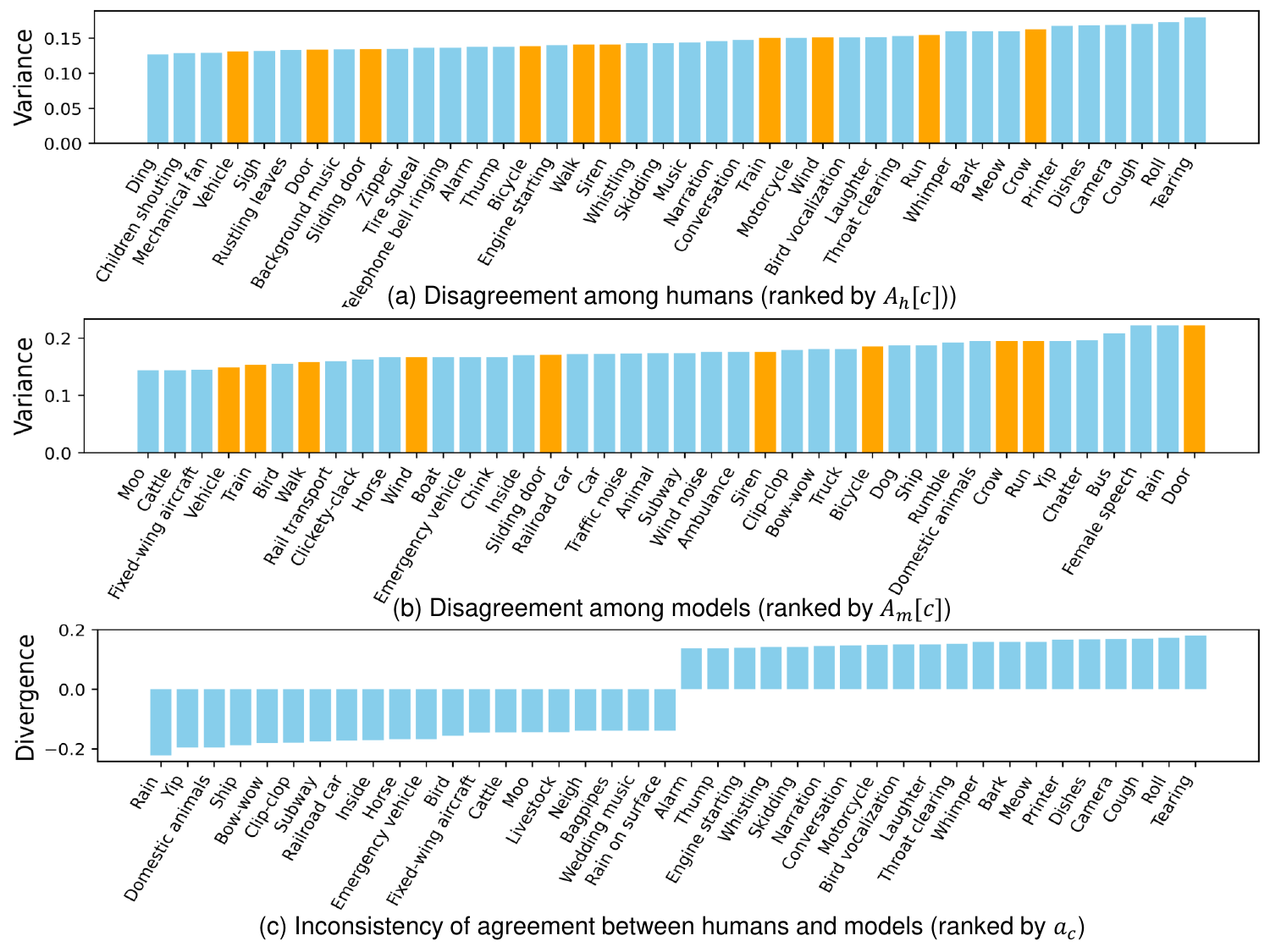}
  \caption{Subfigures (a) and (b) show the top 40 classes that are the most difficult to reach agreement in humans and models, respectively. The orange bars indicate the overlapped classes between the top 40 classes in humans and machines. Subfigure (c) shows the top 40 classes with the largest gap between human perception and model inference, where each $a_c<0$ and $a_c>0$ provide 20 classes.}
  \label{fig:semantic}
\end{figure*}

\item A manual review was still necessary after GPT-4. Labels, particularly for more complex or less common audio events, were reviewed by experimenters, and adjustments were made where required to ensure the labels accurately reflected the audio content.

\item After the initial label mapping by GPT-4 and subsequent manual adjustments, a final consistency check was conducted to ensure uniformity across all annotations. 
\end{enumerate}
At last, the descriptive texts in the MAFAR dataset are aligned to a subset of the AudioSet label set, containing 86 classes. The frequency of aligned classes can be found in Fig. \ref{fig:frequency}.

\section{Benchmark Methods and Resutls}

This work introduces event semantic identification and existence detection to compare the semantic importance in human perception and model inference. The model inference is simulated by an ensemble model group of 6 state-of-the-art (SOTA) pre-trained models on AudioSet \cite{audioset}, including Transformer-based EAT-Base \cite{EAT} and EAT-Large \cite{EAT}, pure convolutional neural network (CNN)-based PANNs-16k \cite{cnn14} and PANNs-8k \cite{cnn14}, as well as two extra models (with mAP = 0.495 and 0.504), which are currently under review and will be publicly available soon. 
For model predictions, this work converts the ensemble of models' output probabilities of audio events to hard labels using a threshold of 0.5, which simulates models' recognition of prominent and noticeable events to comprehensively compare the foreground events labelled by 10 professional annotators in Section \ref{section2}.

\subsection{Difference in Event Semantic Identification}
For the semantic identification of events \cite{atpolyphonic, xu2018large, kim2019comparison, gong2021psla}, humans and models may exhibit different tendencies for specific classes, creating a gap between them\cite{kolajo2023human}. However, a major challenge arises when comparing these identifying tendencies: significant differences can exist between individual human perceptions, and in extreme cases, the variation between two humans may be larger than that between humans and machines. Inspired by Carlini et al. \cite{carlini2019distribution}, which verified the ensemble agreement of models could reflect the representativeness of samples, this work proposes measuring the consistency of agreement between human and model groups to better understand their respective tendencies. 

Specifically, for each audio segment $x_i$, human and model groups will provide an annotated and predicted label set, $H_i=\{y_i^1,y_i^2,\cdots,y_i^s|y_i^s\in R^N\}$ 
and $M_i=\{\hat{y_i^1},\hat{y_i^2},\cdots,\hat{y_i^t}|\hat{y_i^t}\in R^N\}$ respectively, where N is the classes amount. The agreement extent in the humans $A_h$ and models $A_m$ can be further represented by the variance of the corresponding label set:
\begin{align}
    & A_h[c] = \frac{1}{P(c,H)}\sum_{i=P(c,H)} \textit{Var}(H_i)\\
    & A_m[c] = \frac{1}{P(c,M)}\sum_{i=P(c,M)} \textit{Var}(M_i)
\end{align}
where $\textit{Var}(\cdot)$ means the variance calculation, $A[c]$ means the $c$-th class, $P(c,H)$ is an index function to ensure that the variance calculation only involves segments, which may potentially have the events of class $c$, namely $P(c,H) = \{i|sum(H_i)[c] > 0\}$. The $P(c,H)$ is mainly used to prevent the variance from being affected by the frequency of event appearance. For the class $c$, the consistency of agreement extent $a_c$ could be represented as follows:
\begin{align}
    a_c = (A_h-A_m)[c]
\end{align}
where the $a_c > 0$ means that the extent of disagreement in humans is greater than the model, conversely $a_c < 0$. By ranking the $a_c$, it is possible to find those classes with significant gaps between human perception and model inference.

Fig. \ref{fig:semantic} shows the disagreement and gap between human perception and model inference in terms of semantic event identification. In subfigure (a) and (b), comparing with humans and models, the results show that there are 10 same classes in the top 40 classes with the highest group disagreement. The difference is that models are more easily affected by vehicle noise, such as bus, ship, truck, ambulance, subway and Traffic noise, etc., while human seems to find it harder to achieve agreement on subtle, trivial or easily covered events, such as tearing, roll, cough, etc. Intuitively, this result aligns with common sense, as people in cities usually filter out persistent background noise, and different people may pay inconsistent attention to those subtle events \cite{stocker2013hear}. Fig. \ref{fig:semantic} (c) further shows the specific class with significant inconsistency of agreement between humans and models. Compared with models, most events that are harder for humans to reach achievement ($a_c>0$) are consistent with Fig. \ref{fig:semantic} (a). For $a_c<0$, besides the vehicle noise, it is obvious that models are also affected by semantic-similar classes, such as rain and rain on surface, domestic animals, horse, and Moo.

\subsection{Difference in Event Existence Detection}
\begin{figure}[t]
  \centering
  \setlength{\abovecaptionskip}{0.1cm}    
	\setlength{\belowcaptionskip}{-0cm}
  \includegraphics[width=0.4\textwidth, height=4.5cm]{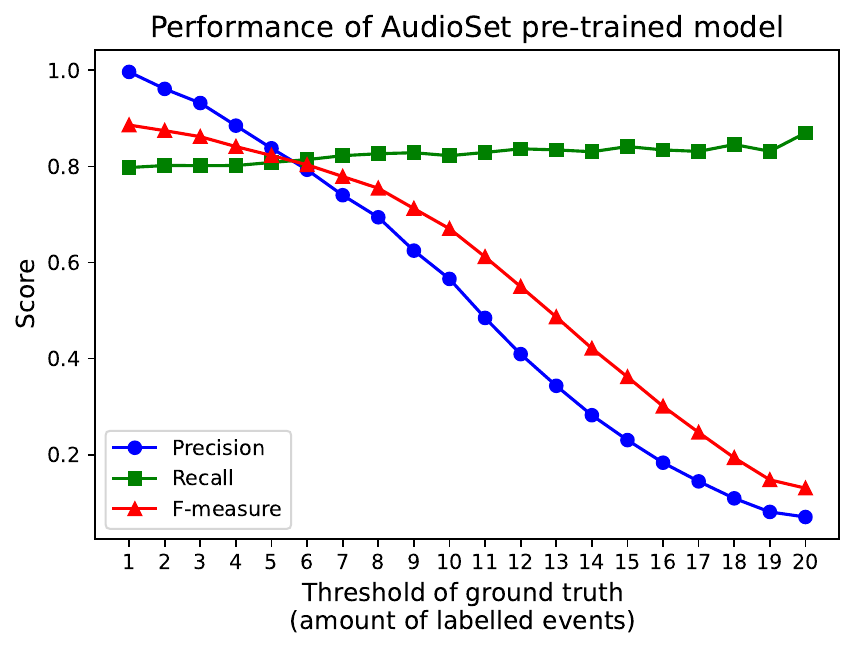}
  \caption{The ensemble results of the existing AudioSet pre-trained models in event existence detection. The ground truth threshold of the existence of interested events is based on the total amount of labelled events from 10 annotators. Since there could be more than one event in segments, the ground truth threshold could exceed 10.}
  \label{fig:existence}
\end{figure}

In addition to semantic identification, some AER tasks \cite{rovetta2020detection,primus2020anomalous,wilkinghoff2021sub,suefusa2020anomalous,wichern2021anomalous} focus solely on detecting the existence of audio events. Existence detection is typically a binary classification applied for data pre-processing or anomalous detection, allowing for greater tolerance of semantic misidentification. The multi-person independent annotations in the MAFAR dataset provide an opportunity to explore the gap between human perception and model inference in event existence detection.

Specifically, the labels from multiple annotators can be used to construct various ``ground truths" regarding the existence of events \cite{cartwright2019crowdsourcing}\cite{humphrey2018openmic}. By counting the number of labels for the same time period, an adjustable threshold can be set to define the ground truth according to the level of human agreement. To assign greater weight to individuals who annotate more events in the same period, this work adopts the total number of labelled events into the threshold judgment process instead of the annotator amount:
\begin{align}
y_{true} = 
\begin{cases} 
    1, & \text{if } \textit{sum}(H_i) \geq \textit{threshold}, \\
    0,  & \text{else}
\end{cases}
\end{align}
where $y_{true} = 1$ represents the existence of the event.

For the AudioSet pre-trained models used in this work, as they are pre-trained multi-class classification tasks, this work uses the maximum value of ensemble prediction of 527 classes as the result of event existence detection:
\begin{align}
y_{pre} = 
\begin{cases} 
    1, & \text{if } \textit{max}(y') > 0.5, \\
    0,  & \text{else}
\end{cases}
\end{align}
where $y'$ is the averaged output of all ensemble models.

Fig. \ref{fig:existence} shows the Precision, Recall and F-measure scores \cite{stowell2015detection} of the ensemble pre-trained models with different ground truth thresholds. With the threshold for the number of human labels increasing, the results show that the Precision and F-measure gradually reduce, while the Recall has a subtle increase. This tendency illustrates that machine inference tends to align with the $\textit{threshold} = 1$, namely that the ground truth is the union set of all labelling results, indicating that models are more sensitive to event existence detection than humans. The Recall curve also indicates that the benchmark tends to ignore 15\%-20\% human-interested events, and this ignoring alleviates slightly as the ground truth threshold increases. 

\section{Discussion}
For the gaps between human perception and current model inference, we suspect that the reason may lie in the labelling sides, where sounds are taken out of context and given way more attention than they would have in real life. Particularly for the AudioSet \cite{audioset}, videos from the YouTube source are uploaded by people who already bias the data towards salient events, as these are typically considered more noteworthy by the public.
Meanwhile, under the existing evaluation criteria of AER (e.g., mAP\cite{audioset}), most methods \cite{cnn14}\cite{EAT} tend to adopt a series of techniques to alleviate the influence of imbalanced data for better performance, while these techniques also may further magnify the salience of some trivial events. Given the current research trends, audio event recognition models may face a dilemma between pursuing better performance metrics and aligning with real human perception, which is worth noting for AER researchers and the DCASE community.

\section{Conclusion}
This work proposes the MAFAR dataset to quantify human perception in AER-related tasks and promote future research on semantic importance in AER systems. By comparing the distribution of labels from multiple annotators, the study captures differences and highlights the gap between human perception and model inference in event semantic identification and existence detection. Experimental results reveal that human perception and model inference have a significant gap in subtle or noisy audio events of semantic identification. In contrast, the ensemble of pre-trained models shows more sensitivity than humans in event existence detection. These gaps may arise from biased datasets and model evaluation criteria, amplifying the salience of certain events. This work calls for more attention to this issue and will continue to annotate more audio clips in different scenes to expand the scale of the MAFAR dataset.


\bibliographystyle{IEEEbib}
\bibliography{refs}

\vspace{12pt}

\end{document}